\documentclass[twocolumn,aps,epsfig,graphics]{revtex4}
\usepackage{graphicx,amsmath}

\def\lsim{\lower.5ex\hbox{$\; \buildrel < \over \sim \;$}}
\def\gsim{\lower.5ex\hbox{$\; \buildrel > \over \sim \;$}}
\def\g{\ifmmode \gamma \else $\gamma$\fi}
\def\gs{\ifmmode \gamma \else $\gamma~$\fi}

\newcommand{\la}{\langle}
\newcommand{\ra}{\rangle}

\begin{document}

\title{Dynamic canonical suppression of strangeness in transport models}

\author{O. Fochler, S.Vogel, M. Bleicher, C. Greiner, P. Koch-Steinheimer, Z. Xu}

\affiliation{Institut f\"ur Theoretische Physik, J.W. Goethe Universit\"at \\
Max-von-Laue-Str. 1, D-60438 Frankfurt am Main, Germany}

\begin{abstract}
It is investigated whether canonical suppression associated with the
exact conservation of an U(1)-charge can be reproduced correctly by
current transport models. Therefore a pion-gas having a volume-limited
production and annihilation cross section for $\pi\pi\leftrightarrow
K\bar{K}$ is simulated within two different transport prescriptions for
realizing the inelastic collisions. It is found that both
models can indeed dynamically account for the canonical suppression in
the yields of rare strange particles.  
\end{abstract}

\maketitle

\section{Introduction}
As the properties of hadronic gases provide important theoretical
means of describing heavy ion collisions, there exist several models
to address this problem. Among the most prominent are statistical
thermal models for equilibrated hadronic gases.

One result is a suppression in the yields for rare
particles when treating the conservation of a corresponding
U(1)-charge exactly \cite{Redlich:2001it,Rafelski:1980gk,Ko:2000vp}. The
importance of such a treatment is fortified by recent experimental
results \cite{Oeschler:2000dp}. The suppression effect can eventually be formulated in terms of a
volume-dependence \cite{Rafelski:1980gk}.

A dynamical interpretation has recently been offered by the formulation and solution of
kinetic master-equations, which are capable of covering the time
evolution of the system as well \cite{Ko:2000vp}. 

In the following we will demonstrate that standard hadronic transport
models dynamically show canonical suppression. More
explicitly the U(1)-charge considered is the strangeness 
whose net value is taken to be zero throughout our calculations. Due
to simplicity we consider only inelastic reactions of the type
$\pi\pi\leftrightarrow K\bar{K}$, where the kaons and anti-kaons bear 
strangeness $+1$ and $-1$ respectively.  

An abundant number of kaons justifies a grand canonical description of
the system, in which the strangeness is conserved only on the average $\langle N_{K} \rangle - \langle
N_{\bar{K}} \rangle = 0$. The density of particles having strangeness
$+1$, the kaons, can then be computed
\begin{equation} \label{gcdens}
n^{\mathrm{gc}}_{K}=\frac{Z_{K}^{1}}{V},
\end{equation}
where $Z_{K}^{1}$ denotes the relativistic one-particle partion function for
non-interacting particles (kaons in this case)
\begin{equation} \label{1partZ}
Z^{1}_{K}\left(V,T\right)=g_{K}\frac{VT}{2\pi^{2}}m_{K}^{2}K_{2}\left(\frac{m_{K}}{T}\right).
\end{equation}

A rare number of kaons however demands the conservation of strangeness
to be treated exactly, i.e. $N_{K}-N_{\bar{K}}=0$. A consistent realization of this
constraint eventually leads to a canonical partition function describing the
system \cite{Rafelski:1980gk,Redlich:2001it} and thus to the following kaon density
\begin{align} \label{cdens}
n^{\mathrm{c}}_{K}= \eta\ n_{K}^{gc},
\end{align}
with the canonical suppression factor $0\leq\eta\leq1$ being
\begin{equation} \label{eta}
\eta=\frac{n_{K}^{\mathrm{c}}}{n_{K}^{\mathrm{gc}}}=\frac{I_{1}(x)}{I_{0}(x)}
\end{equation} 
and $x=2Z_{K}^{1}$. Pursuing the kinetic master equation approach
yields the same results for the equilibrated kaon densities \cite{Ko:2000vp}.

Investigating the canonical suppression factor (\ref{eta}) more
closely, using the asymptotic behaviours
\begin{align}\label{asymptI}
\lim_{x\to\infty}\frac{I_{1}(x)}{I_{0}(x)}&\rightarrow 1&
\lim_{x\to 0}\frac{I_{1}(x)}{I_{0}(x)}&\rightarrow \frac{x}{2},
\end{align}
one finds that in the grand canonical limit the kaon density $n_{K}$
is independent of the reaction volume, whereas in the canonical
regime, with the number of kaons $\la N_{K}\ra\ll 1$,
it scales linearly with the volume as $x\propto V$. This volume
dependent behaviour of the kaon density (\ref{cdens}) provides a
convenient reference for comparison between simulation and 
theory.

\section{Dynamical simulations}
The simulation setup consists of a large box of $20\ \mathrm{fm}$
side length holding a relativistic gas of pions. This pion-gas provides
a heat bath for a much smaller reaction volume of variable size centered within
the large box. Inside this likewise box-shaped reaction volume
processes $\pi\pi\leftrightarrow K\bar{K}$ are allowed, covering all
possible isospin states of pions and kaons. The kaons are
reflected by the walls of the reaction volume,  whereas these walls are permeable for the
pions. After equilibration, the kaon density within the reaction volume
should be governed by (\ref{cdens}) which holds as a reference for the
transport model results. Therefore different sizes of the inner
reaction volume are simulated and the number of
kaons $\la N_{K}\ra$ is extracted by averaging over many timesteps. The
timesteps are sufficiently large such that no correlations are to be expected.

We implemented this scenario using two different types of transport
descriptions - the microscopic transport model UrQMD
\cite{Bass:1998ca} and a realization of a stochastic
transport model borrowed from a recent and elaborated parton cascade
\cite{Xu:2004mz}. The former model makes use of a geometrical
interpretation of cross sections in order to solve transport
equations, whereas the latter relies on the calculation of 
transition probabilities. 

The UrQMD model provides full space time dynamics for hadrons and
strings. It is a non-equilibrium model based on the covariant
propagation of hadrons and strings and the geometric interpretation of
cross sections. For our studies it was modified such that only
reactions $\pi\pi\leftrightarrow K\bar{K}$ together with all possible elastic
$2\leftrightarrow 2$ processes remained possible. We had to choose sufficiently small cross
sections such that the mean free path remained large compared to the interaction length
$\lambda_{\mathrm{m.f.p.}}=(n\sigma_{22})^{-1}\gg\sqrt{\sigma_{22}/\pi}$.
Otherwise the difference in the collision times of the involved
particles viewed from the lab frame causes a decrease in the
collision rates as pointed out in \cite{Xu:2004mz}. As different
densities are involved, forward and reverse reactions are not affected in the same way
and the change in reaction rates leads to a shifted equilibrium value
of the kaon density. For details see \cite{Fochler:2005inprep}.

The simulation of $2\leftrightarrow 2$ processes within the stochastic
method is based on the calculation of a collision probability for each
possible particle pair per unit volume $\Delta x^{3}$ and unit time $\Delta t$ via
\begin{equation} \label{collProb}
P_{22}=v_{rel}\sigma_{22}\frac{\Delta t}{\Delta x^{3}}.
\end{equation}
$v_{rel}$ denotes the relative velocity and $\sigma_{22}$ is the cross
section for the considered $2\leftrightarrow 2$ process. Any so obtained
probability is then compared with a random number
between $0$ and $1$ to decide whether the collision should take place
or not.

The initial conditions in any model are chosen such that the pion
gas acquires a temperature of~$T=170\ \mathrm{MeV}$. The appropriate
number of pions and the total energy of the system are calculated via
the use of a grand canonical partition function for pions
alone. Our heat bath volume of $8000\ \mathrm{fm}^{3}$ corresponds to
a population of $1348$ pions bearing a total energy of $747.5\
\mathrm{GeV}$. Initially each pion is assigned the same fraction of
the total energy, giving one half of the particles momenta in the
positive $x$-direction, while the remaining particles start out
bearing momenta in the negative $x$-direction. The spatial 
distribution is random.

\begin{figure} [phbt]
\centering
\includegraphics[width=.5\textwidth,angle=0]{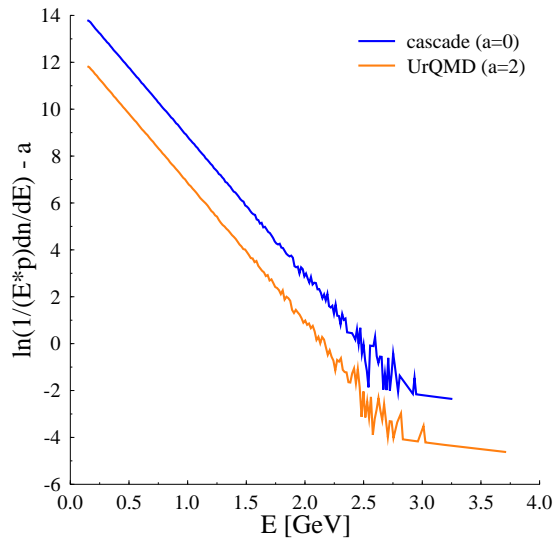}
\caption{Logarithmic energy spectra for the simulated
pion-gases.}
\label{fig_spectrum}
\end{figure}

A fundamental and necessary verification of the simulations'
reliability is of course a check for
thermalization. Figure~\ref{fig_spectrum} nicely demonstrates that the
energy distribution of the pions becomes thermal in both models.

\begin{figure} [phbt]
\centering
\includegraphics[width=.5\textwidth,angle=0]{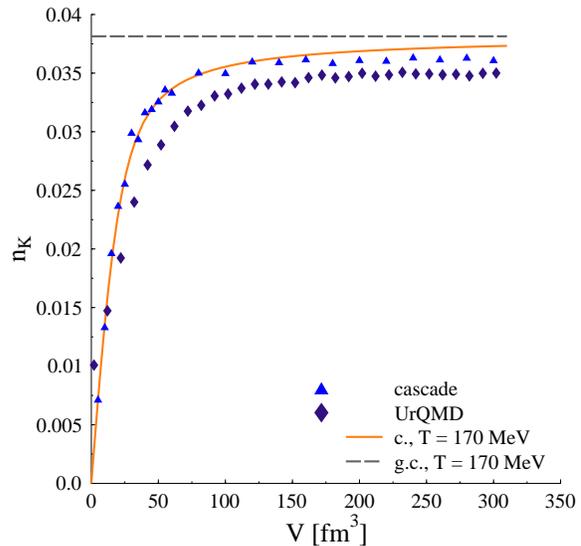}
\caption{Kaons density versus reaction volume as extracted from
simulations with UrQMD (diamonds) and the stochastic cascade
(triangles). For comparison the dashed line indicates the grand
canonical behaviour. The solid line shows the canonical volume dependence of the kaon
density as expected from (\ref{cdens}) for a temperature of $T = 170\ \mathrm{MeV}$.}
\label{fig_densSim}
\end{figure}

Figure~\ref{fig_densSim} then displays our results in terms of
the extracted kaon density plotted versus the reaction volume along
with the theoretical predictions for the grand canonical
(\ref{gcdens}) and canonical (\ref{cdens}) behaviour. The minor
fluctuations in the results provide a visual indication for the
irrelevance of statistical errors. It is manifest that canonical
suppression is reproduced by both models, i.e., as expected the kaon 
yield is suppressed with respect to the grand canonical limit for
small reaction volumes and thus small numbers of produced kaons.    

The canonical suppression for the number of kaons $\la N_{K}\ra$ being
considerably smaller than $1$, as pointed out in \cite{Ko:2000vp},
originates dynamically from an enhancement of the annihilation process
by $1/\la N_{K}\ra$ as compared to standard, grand canonical
formulation of the Boltzmann equation. The reason is that any kaon in
the system requires the existence of a corresponding
anti-kaon due to strangeness conservation. Thus the probability of
finding a particle anti-particle pair turns into a highly correlated conditional
probability. The so enhanced annihilation probability then leads to
the canonical suppression in the kaon yields. As shown by our results, this
behaviour is automatically included in the considered transport models.


\end{document}